\documentclass[%
superscriptaddress,
reprint,
amsmath,amssymb,
aps,
prl,
]{revtex4-1}

\usepackage{graphicx}
\usepackage{dcolumn}
\usepackage{bm}
\usepackage{hyperref}

\begin{document}

\title{Persistence and lifelong fidelity of phase singularities in optical random waves}

\author{L. De Angelis}
\affiliation{%
	Kavli Institute of Nanoscience, Delft University of Technology, 2600 GA, Delft, The Netherlands
}%
\affiliation{%
	Center for Nanophotonics, AMOLF, Science Park 104 1098 XG Amsterdam, The Netherlands
}%
\author{F. Alpeggiani}
\affiliation{%
	Kavli Institute of Nanoscience, Delft University of Technology, 2600 GA, Delft, The Netherlands
}%
\affiliation{%
	Center for Nanophotonics, AMOLF, Science Park 104 1098 XG Amsterdam, The Netherlands
}%
\author{A. Di Falco}
\affiliation{
	SUPA, School of Physics and Astronomy, University of St Andrews, North Haugh, St Andrews KY16 9SS, UK
}
\author{L. Kuipers}
\email{l.kuipers@tudelft.nl}
\affiliation{%
	Kavli Institute of Nanoscience, Delft University of Technology, 2600 GA, Delft, The Netherlands
}%
\affiliation{%
	Center for Nanophotonics, AMOLF, Science Park 104 1098 XG Amsterdam, The Netherlands
}%

\date{\today}

\begin{abstract}
Phase singularities are locations where light is twisted like a corkscrew, with positive or negative topological charge, depending on the twisting direction. Among the multitude of singularities arising in random wave fields, some of them can be found at the same location, but only when they exhibit opposite topological charge, which results in their mutual annihilation. New pairs can be created as well. With near-field experiments supported by theory and numerical simulations we study persistence and pairing statistics of phase singularities in random optical
fields as a function of the excitation wavelength. We demonstrate
how such entities can encrypt fundamental properties of the random fields in which they arise.
\end{abstract}

\pacs{Valid PACS appear here}
\maketitle

A wide variety of physical systems
exhibit vortices: locations around which an observable rotates
while being undetermined in the
middle~\cite{Zwierlein2005,Lagoudakis2008,Auslaender2009,Roumpos2011,Manni2012,Giomi2015,Barboza2015,Barnett2017,Souslov2017}.
It is exceptionally fascinating when the
properties and evolution of such
singular entities can comprehensively describe complex phenomena
such as the Kosterlitz-Thouless transition~\cite{Kosterlitz1973}.
But vortices are not a peculiarity of superconductors:
light's phase swirls around optical vortices, where it is singular~\cite{Nye1974}.
A multitude of these phase singularities arises in
random optical fields, one half swirling in opposite direction
to the other, so that they can approach respectively to an
arbitrarily small distance~\cite{Berry2000,OHolleran2009,Stockmann2009,DeAngelis2016,Taylor2016}.
It is by letting them move that one can observe
creation and/or annihilation of such pairs~\cite{Rozas1997,OHolleran2008,Dennis2010,Giomi2013,
	Cheng2014,Rotenberg2015}.

With near-field experiments we track phase singularities in a
random optical field from \emph{birth} to \emph{death}.
We map the singularities' trajectories at varying the
excitation wavelength, and quantitatively
determine properties such as their persistence in the field and
the correspondence between creation and annihilation partner of a singularity,
known as \emph{lifelong fidelity}~\cite{Berry2015}.
We observe two populations of singularities, neatly differentiated
by their typical persistence in the varying wave field:
short-lived pairs that are predominantly \emph{faithful}
to their creation partner, and a more promiscuous population.

\begin{figure*}[t]
	\centering
	\includegraphics[width = \textwidth]{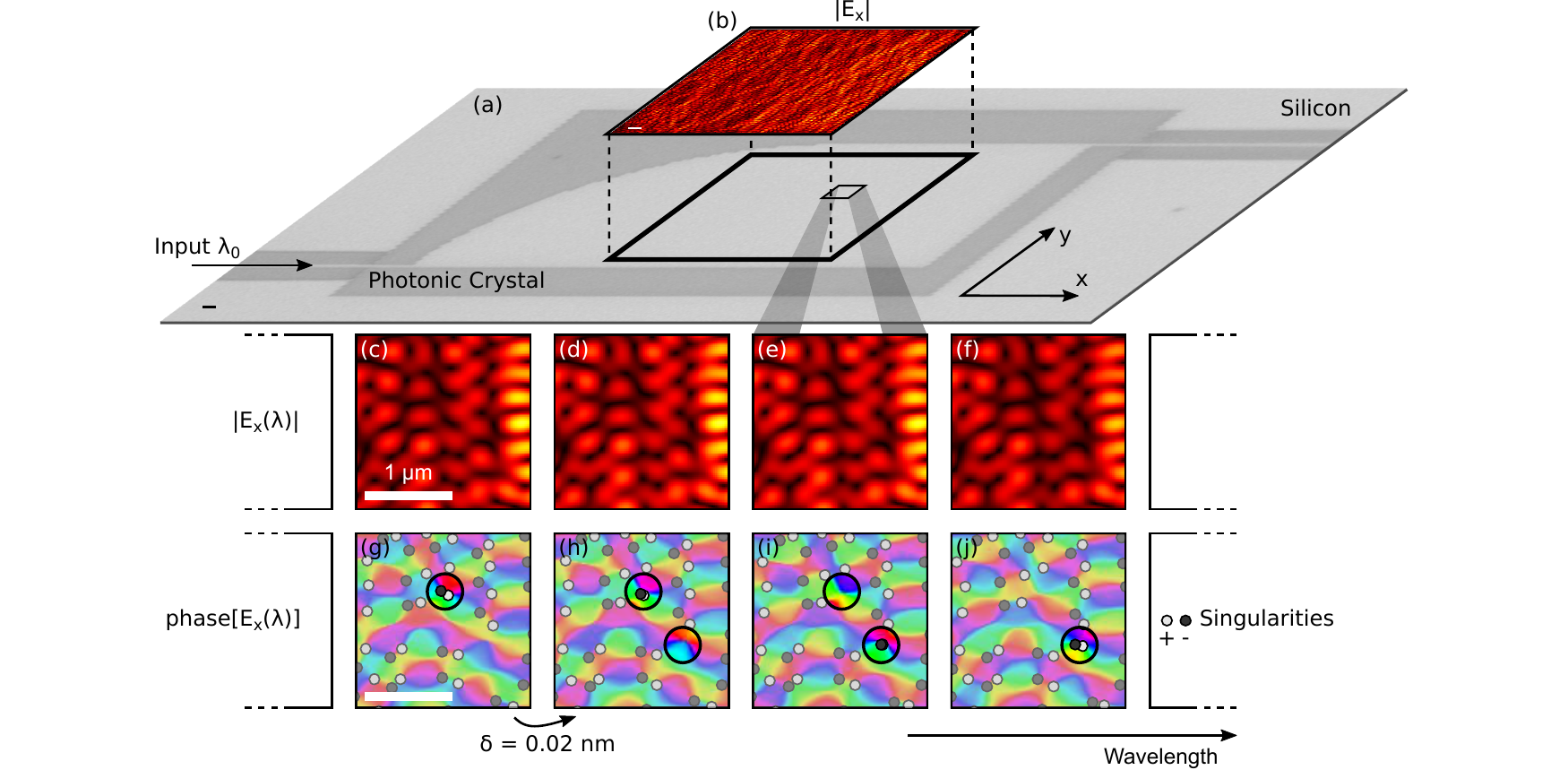}
	\caption{Near-field measurements of the optical field in
		the chaotic cavity. (a) Optical micro-graph of the chaotic cavity: 
		the dark area is a photonic crystal that confines light inside the cavity.
		(b) $17\,\mathrm{\mu m }\times 17\,\mathrm{\mu m }$ near-field 
		map of the amplitude of the $x$ component of the electric
		field in the cavity.  
		(c-f) Zoom-in of panel b for different
		input wavelength $\lambda$. (g-j) False-color
		map of the measured phase for different input wavelength
		$\lambda$; in such maps we pinpoint phase singularities with positive
		(dark spots) or negative (light spots) topological charge. The
		black circles highlight annihilation and creation events.}
	\label{fig:meas}
\end{figure*}
To generate optical random waves we couple monochromatic light
($\lambda_0 \sim 1550\,\mathrm{nm}$) into a photonic crystal chaotic 
cavity realized in a silicon-on-insulator platform (Fig.~\ref{fig:meas}a)~\cite{Liu2015}.
We map the optical field
inside the cavity with near-field microscopy,
resolving amplitude, phase and polarization of such in-plane complex electric field
$(E_x,E_y)$~\cite{balistreri2000,Burresi2009,leFeber2014}.
Figure \ref{fig:meas} presents a typical example of our
measurements of amplitude and phase of~$E_x$.
The optical field inside the cavity
can be thought of as a random superposition of light waves~\cite{Stockmann2006}
with transverse electric (TE) polarization~\cite{DeAngelis2016}.
Only the behavior of~$E_x$ is presented here, without loss of
generality as it is representative of the behavior of all in-plane
field components~\cite{DeAngelis2016}.

Figure~\ref{fig:meas}b represents a full-size measurement:
a square map
$17\,\mathrm{\mu m }\times 17\,\mathrm{\mu m }$
with a pixel size of about $17\,\mathrm{nm}$.
In this map we distinguish
a multitude of \emph{dark}
and \emph{bright} spots,
the results of destructive and constructive interference.
Figures~\ref{fig:meas}c-f are $2\,\mathrm{\mu m }\times 2\,\mathrm{\mu m }$
zoom-in of the full measurements taken at different wavelengths of
the input light. Here, we can observe how a small change of the wavelength
($\delta = 0.02\,\mathrm{nm}$), hardly changes the spatial pattern of the
amplitude.
Figures~\ref{fig:meas}g-j depict the measured phase, and
reveal the phase singularities which take place at every zero
in the amplitude (gray circles). 
We pinpoint the position of these singularities
with deep sub-wavelength resolution,
and simultaneously determine their topological charge, which is always
observed to be +1 (light circles) or -1 (dark circles),
corresponding to a $\pm 2\pi$ change of the phase around the singular
point~\cite{Gbur16}.

\begin{figure}[b]
	\centering
	\includegraphics{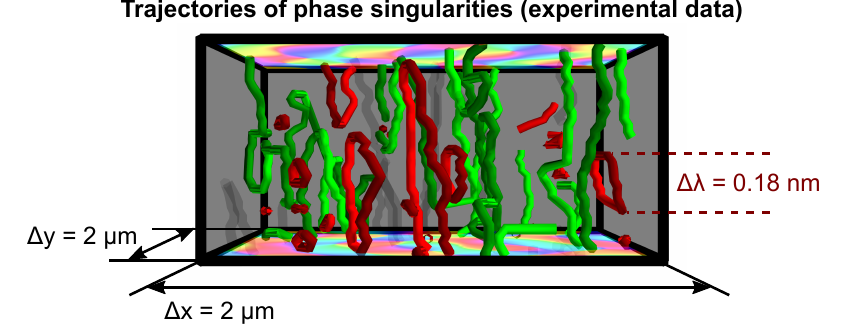}
    \caption{3D representation of the trajectories
    of singularities that propagate through
    a $2\,\mathrm{\mu m }\times 2\,\mathrm{\mu m }\times 0.64\,\mathrm{nm }$
    observation volume. The red and green trajectories are of
    faithful and unfaithful singularities, respectively. In both cases bright
    and dark colors differentiate oppositely charged singularities.
    Please note that the parts of the
    trajectories that continue outside of the observation volume
    are not shown.
    The semi-transparent trajectories are of singularities that propagate
	outside of the total wavelength range: these are not taken into account
	in our statistics.}
    \label{fig:trajectories}
\end{figure}

In the panels g-j of Fig.~\ref{fig:meas}, we observe that the
singularities move as a function
of wavelength.
More eye-catching than their tiny movements are
the annihilation and creation
events of pairs of singularities,
which we can both observe between the panels h and i (Fig.~\ref{fig:meas}),
highlighted by black circles.
Indeed, singularities can be created and annihilated, but only in
processes that conserve the total topological charge of
the system, i.e., in pairs of opposite topological charge.
Thus,
as we tune the wavelength, singularities exhibit a transitory
persistence in the random field over the span of a finite wavelength
shift $\Delta\lambda$ between their creation and annihilation. 
In Fig.~\ref{fig:trajectories} we present a 3D representation of the trajectories
followed in space and wavelength by a small subset of all the singularities that we
measure.
Among all the trajectories, the red ones represent \emph{lifelong faithful}
singularities:
\emph{special} cases where the singularities have
the same partner for both creation
and annihilation. The green trajectories are \emph{unfaithful} singularities.

\begin{figure*}[t]
	\centering
	\includegraphics[width = \textwidth]{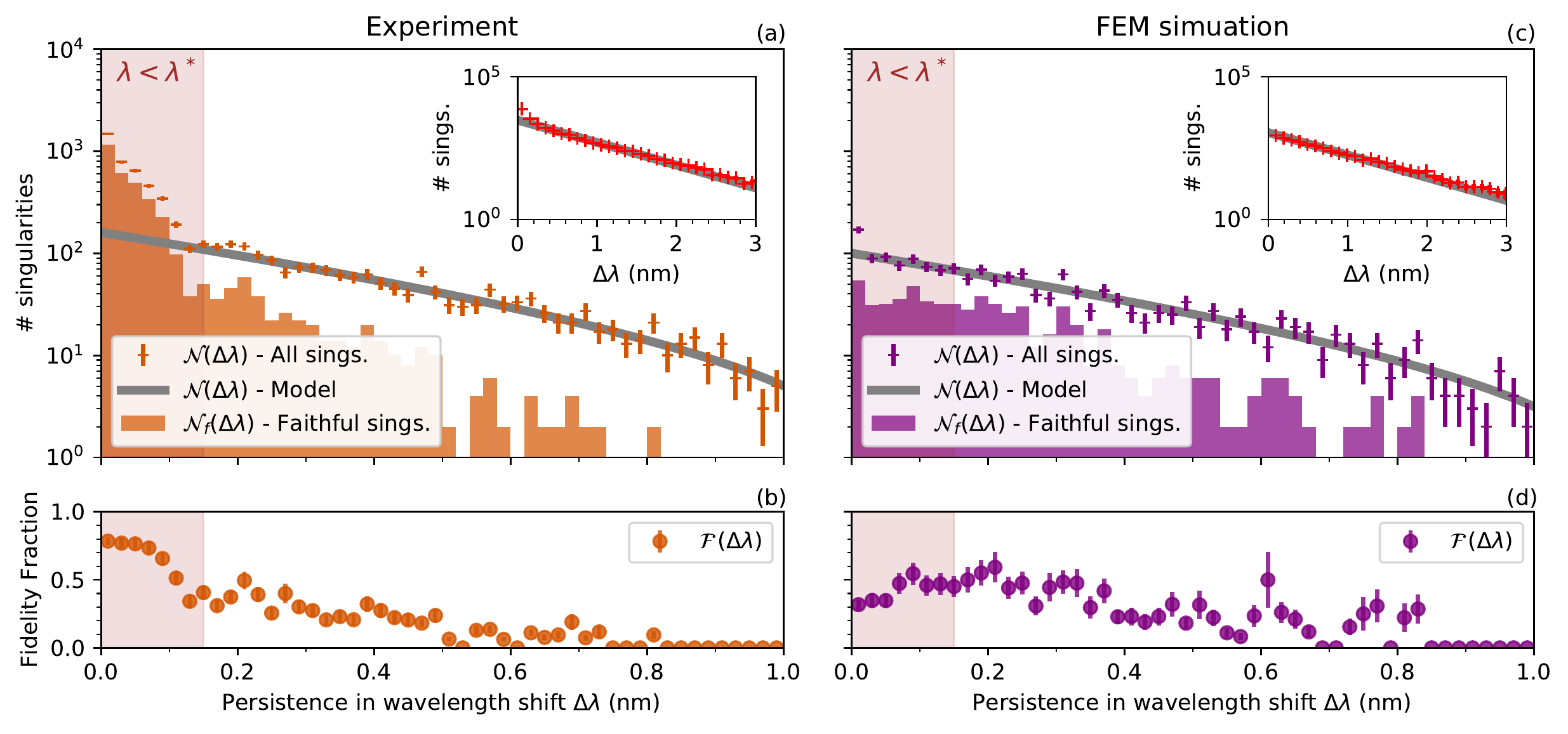}
	\caption{Overview of the statistics for the persistence and fidelity
		of singularities in random waves, in experiment (a,b) and FEM
		simulation (c,d). In the upper panels the dashes indicate
		the number of singularities  $\mathcal{N}$ which persisted
		in the field
		for a given wavelength shift $\Delta\lambda$. The main figures
		refer to experiment/simulation where the wavelength was sweeped over
		a range $\mathcal{L}=1.2\,\mathrm{nm}$
		with a step $\delta = 0.02\,\mathrm{nm}$, whereas the
		insets show the results of wider scans ($\mathcal{L}=8\,\mathrm{nm}$,
		$\delta = 0.1\,\mathrm{nm}$).
		The gray lines are representative of the prediction
		for the persistence histogram given by our model.
		In the same panels, the boxes are still a persistence histogram, but
		in which only the $\mathcal{N}_f$ singularities
		that were faithful to each other are taken into
		account.
		In the lower panels we report the fidelity fraction $\mathcal{F}$
		of phase singularities, again as a function of their persistence
		in the field [$\mathcal{F}(\Delta\lambda) =
        \mathcal{N}_f(\Delta\lambda)/\mathcal{N}(\Delta\lambda)$].}
	\label{fig:persistences}
\end{figure*}
Figure~\ref{fig:persistences}a presents the number $\mathcal{N}$
of singularities as a function of their persistence $\Delta\lambda$
in the measured field.
The main plot illustrates the results obtained from
a dataset in which
we tuned the wavelength with a step 
$\delta = 0.02\,\mathrm{nm}$
for a total range $\mathcal{L} = 1.2\,\mathrm{nm}$.
In such plot we clearly distinguish two
persistence regimes separated by a cutoff wavelength shift
$\lambda^*\approx 0.15\,\mathrm{nm}$.
In the region $\Delta\lambda > \lambda^*$ the number of
singularities exponentially decays against their
persistence in the field,
with a characteristic persistence $\lambda_d = 0.6\,\mathrm{nm}$.
Such exponential behavior is even more clear from the
data displayed in the inset, representative of
a measurement
in which the total range of the wavelength sweep
is $\mathcal{L} = 8\,\mathrm{nm}$ ($\delta = 0.1\,\mathrm{nm}$).
Please note that
the finite size of the wavelength scans has non-negligible
effects on the
measured statistics, mainly for the
longest trajectories, thus slightly
distorting the exponential behavior.
This is because singularities
that are created and/or annihilated
outside the measured wavelength range
need to be excluded
from the persistence statistics.
We estimate
the fraction of such singularities to be of the order
of $\Delta\lambda/\mathcal{L}$,
resulting in a correction factor for 
$\mathcal{N}(\Delta\lambda)$ proportional to
$(1-\Delta\lambda/\mathcal{L})$.

The exponential behavior that we observe could
be interpreted in a straightforward way 
as the result of a memoryless Poisson process.
However, we discover
a physics richer than that.
From the main plot of
Fig.~\ref{fig:persistences}a
we notice that $\mathcal{N}(\Delta\lambda)$
is not purely exponential: it clearly deviates
from such a distribution for $\Delta\lambda$
smaller than $\lambda^*$.
This spectral region contains an excess of singularities
compared to what the asymptotic exponential distribution
would predict. Their characteristic persistence in the field
is much smaller than $\lambda_d$.
In fact, by fitting
$\mathcal{N}(\Delta\lambda)$ with a bi-exponential distribution,
we can estimate the characteristic persistence
for such short-living population
to be approximately $0.03\,\mathrm{nm}$.
Please note that the cutoff wavelength shift $\lambda^*$
does not depend on the absolute starting
wavelength as the trajectories of the singularities
in excess are uniformly distributed along
the measured wavelength range.”

Interestingly, when considering only the
\emph{faithful} singularities
(red trajectories in Fig.~\ref{fig:trajectories}), 
a different behavior is observed than for the full ensemble of all singularities.
It is clear from Fig.~\ref{fig:persistences}a that
in the region where deviation from a single exponential
takes place
($\Delta\lambda < \lambda^*$), we have an over-representation
of faithful singularities (yellow boxes).
This is reflected explicitly
in the fidelity fraction
$\mathcal{F}(\Delta\lambda)=\mathcal{N}_f(\Delta\lambda)/\mathcal{N}(\Delta\lambda)$ 
represented in Fig.~\ref{fig:persistences}b:
while a majority of the \emph{short-living} singularities is
faithful to each other, 
the opposite is true for \emph{long-living} ones.

\begin{figure}[b]
	\centering
	\includegraphics[width = 3.15in]{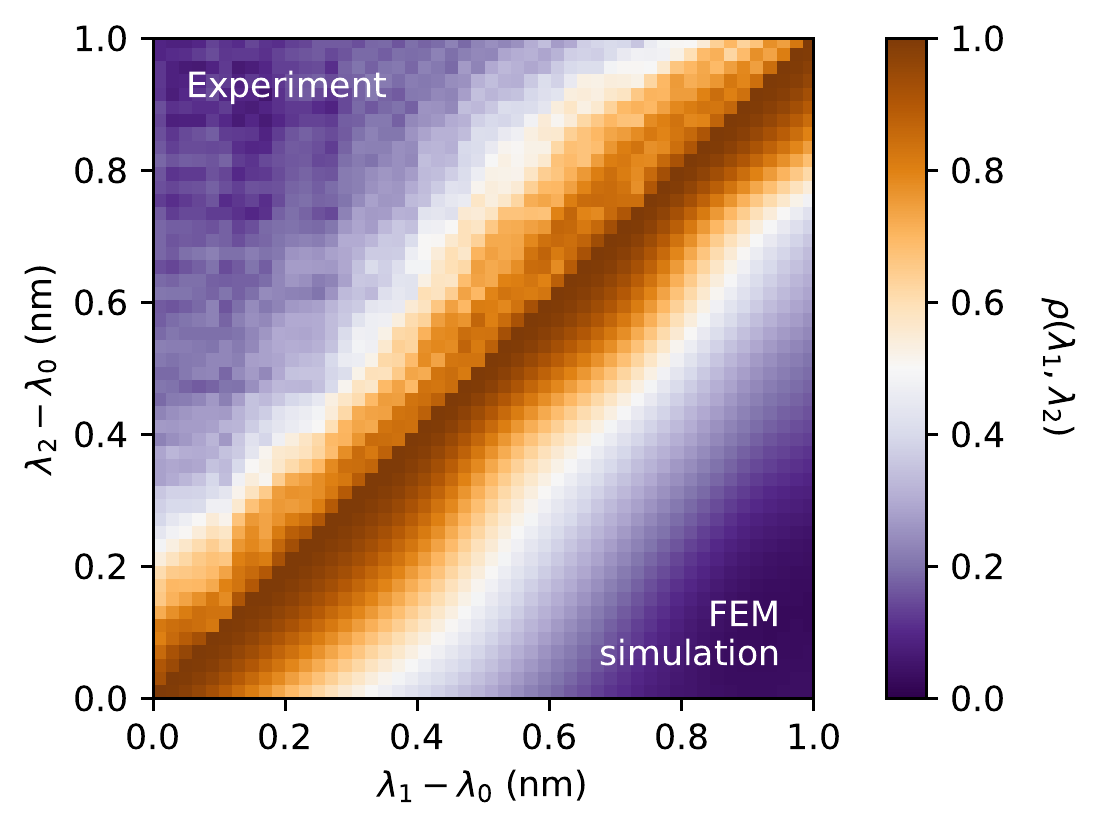}
	\caption{Correlation coefficient
		$\rho(\lambda_1,\lambda_2)$
		of the measured (upper left) and FEM simulated (lower right) optical random field, at wavelengths
		$\lambda_1$ and $\lambda_2$ ($\lambda_0 = 1550\,\mathrm{nm}$).}
	\label{fig:corr}
\end{figure}

The origin of the cutoff $\lambda^*$ which discriminates the population of
faithful and short-living singularities from that of unfaithful and long-living ones
must be sought in the evolution properties of the field.
Figure~\ref{fig:corr}a displays
the correlation coefficient of the considered experimental field
at wavelengths $\lambda_1$ and $\lambda_2$:
\begin{equation}
\rho(\lambda_1,\lambda_2) 
=
\frac{| \int\mathrm{d}\mathbf{r}\,{\tilde{E}_x}^*(\lambda_1)\tilde{E}_x(\lambda_2)|}
{
	\sqrt{\int\mathrm{d}\mathbf{r}\, |{\tilde{E}}_x(\lambda_1)|^2 \,
		\int\mathrm{d}\mathbf{r}\, |{\tilde{E}}_x(\lambda_2)|^2}}
\label{eq:corr},
\end{equation}
where $\tilde{E}_x = E_x -
\langle E_x \rangle$. It is interesting to note that
the correlation coefficient 
$\rho(\lambda_1,\lambda_2)$ decays over a finite wavelength shift
$\lambda_c \approx \lambda^*$.
Such a close relation between $\lambda_c$ and $\lambda^*$
may suggest that those singularities which spend their entire existence
in the region of spectral correlation of the field
exhibit persistence and pairing properties that are different 
from those of singularities over-living this region.

Although seemingly intuitive, this simple
interpretation cannot be the whole story.
We can firmly state this, having developed
a model that creates a field with correlation properties analogous to those
of our measured field, which in fact
does not contain any excess of \emph{faithful}
singularities for small persistences.

We model the frequency-dependent field
as a superposition of the cavity eigenstates $\mathcal{E}_x^j$, centered 
at frequencies $\tilde{\omega}_j = \omega_j + i\gamma_j$~\cite{Stockmann2006}:
\begin{equation}
E_x(\mathbf{r},\omega) = \sum_j\frac{\alpha_j\mathcal{E}_x^j(\mathbf{r},\omega_j)}{\omega - (\omega_j + i \gamma_j)}.
\label{eq:eigendec}
\end{equation}
Based on the size of our cavity we can estimate the
average spectral separation between two consecutive
eigenstates $\Delta = \langle \omega_{j+1} - \omega_j\rangle_j$
to be approximately $0.08\,\mathrm{nm}$~\cite{Stockmann2006}.
In the simplest model field, we assume the eigenstates
to be equidistant in frequency (spacing $\Delta$), with a constant width~$\gamma$
and a unitary weight ($\alpha_j = 1$).
We set $\gamma= 0.16\,\mathrm{nm}$, equal to 
the average loss rate of our system,
which we determine from a
finite difference time domain (FDTD)
simulation of the entire three-dimensional chaotic cavity.
Finally, following Berry's
hypothesis~\cite{Berry1977},
we consider every eigenstate to be a random
superposition of monochromatic plane
waves:
\begin{equation}
\mathcal{E}_x^j(\mathbf{r},\omega_j) = \sum_{|\mathbf{k}| = n\omega_j/c}
a_{x,\mathbf{k}}\,
\exp(i \mathbf{k}\cdot\mathbf{r} + i \delta_{\mathbf{k}}),\label{eq:rwsuperpos}
\end{equation}
where $\delta_{\mathbf{k}}$ is a random variable uniformly distributed
in~$[0,2\pi]$ and
$a_{x,\mathbf{k}}\propto|\mathbf{k}\times\hat{x}|$
is the polarization coefficient for the TE mode~\cite{DeAngelis2016}. 
With these assumptions, we construct the wavelength-dependent
field of Eq.~\eqref{eq:eigendec} and
determine the statistics of its singularities.
We find that the persistence of singularities is
exponentially distributed, with a characteristic
decay rate that
depends on the ratio between the
eigenstates' width and spacing $\gamma/\Delta$.
For our estimated
parameters ($\gamma/\Delta \approx 2$)
this results in $\lambda_d^{th} = 0.6\,\mathrm{nm}$.
The persistence calculated with our model is
presented in Fig.~\ref{fig:persistences}a
(gray line). In the region $\Delta\lambda > \lambda^*$
we obtain perfect matching between experiment and theory.

These theoretical results are confirmed by a 2D finite element
method (FEM) simulation.
The FEM simulation produces a direct expression of
$\mathbf{E}(\mathbf{r},\omega)$,
free of the previous model assumptions.
The statistics for the singularities
obtained from these simulations
are reported in Fig.~\ref{fig:persistences}c-d.
Comparing the simulated persistence histogram 
to our model calculation we notice perfect
agreement at every wavelength shift. 
No deviation from
a single-exponential distribution in the region
$\Delta\lambda < \lambda^*$ is present, and in the same region
the fidelity fraction is lower than
exeperimentally observed (Fig.~\ref{fig:persistences}b,d).
It is interesting that we find a close correspondence between
the correlation properties of measured and
simulated fields: in both cases the correlation coefficient 
$\rho(\lambda_1,\lambda_2)$ decays over a finite wavelength shift of
(Fig.~\ref{fig:corr}), measurable in a half-width-half-max value of
$0.22(3)\,\mathrm{nm}$ for the experiment and $0.26(3)\,\mathrm{nm}$ for the simulation.
We can therefore exclude the different
behavior of the \emph{short-living} population of singularities
to originate merely as a consequence of the finite
spectral correlation of the random wave field.

We note that we ruled out the eventuality
in which the \emph{short-living}
population of singularities is generated by
experimental artifacts such as
noise in the measurements, temperature fluctuations
or phase drifts. With this regard, independent measurements realized with
different wavelength sweeps showed perfect consistency, demonstrating that
variables which are not the wavelength shift cannot affect the final results~\cite{supplemental}.
Moreover,
we introduced these as well as other possible 
measurement artifacts
(i.e., perturbation from the near-field probe,
frequency instabilities)
in our models, in order to check if they could in any case 
lead to an enhanced population of faithful singularities at
small $\Delta\lambda$: they did not show any.

\begin{figure}[h]
	\centering
	\includegraphics[width = \linewidth]{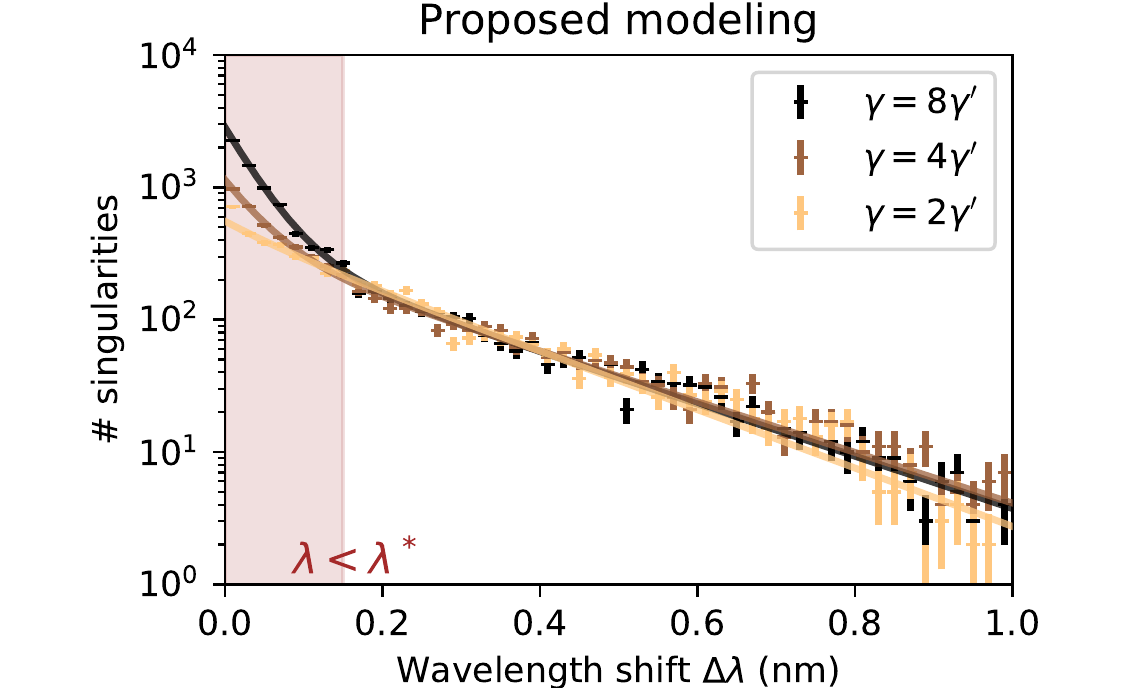}
	\caption{Persistence histograms for singularities arising in numerically generated
		random fields in which two families of eigenstates with different spectral width
		$\gamma$ and $\gamma'$ coexist~\cite{supplemental}. The lines are bi-exponential fits
		to the data points.}
	\label{fig:new_model}
\end{figure}

Interestingly, we did find a modeling that reproduces the observed
enhanced population.
When a second family of eigenstates with
a spectral width $\gamma'$ different from the one
of the original eigenstates
is added to the decomposition of
Eq.~\eqref{eq:eigendec},
two populations of singularities start to appear.
Such an additional family of modes
could be provided by resonances which due to polarization or physical separation
would ideally remain orthogonal
to the chaotic modes, but which get coupled to them in the real-life system.
Figure~\ref{fig:new_model} presents the persistence histogram for such
proposed model case, at varying the ratio $\gamma/\gamma'$. 
For some choices of $\gamma'$ a bi-exponential behavior appears.
Specifically, we observed that such choices satisfy $\Delta\simeq\gamma\gg\gamma'$,
in which cases an excess of faithful singularities living within
the spectral correlation region of the field is again found~\cite{supplemental}.

To conclude, we studied the persistence and pairing statistics
of phase singularities in optical random waves.
For singularities with a persistence
longer than the spectral correlation of the random field,
we find perfect agreement between experiment,
simulation and theory.
It is striking that, for singularities
with a persistence that falls within the spectral
correlation of the random field, we experimentally
observe an excess of singularities compared to theoretical prediction, and
these are more faithful than expected.
With this regard, we propose a mechanism
based on the coexistence of different families of eigenstates,
which could lead to a full
explanation of our experimental observation.

\begin{acknowledgments}
We thank Thomas Bauer for critical readings of the manuscript.
This work is part of the research
program of the Netherlands Organization for Scientific Research (NWO).
The authors acknowledge
funding from the European Research Council (ERC Advanced Grant
No. 340438-CONSTANS). F.A. acknowledges support from the Marie Sk\l{}odowska-Curie individual fellowship BISTRO-LIGHT (No. 748950).
A.D.F. acknowledges support from EPSRC (Grants No. EP/L017008/1 and EP/M000869/1).
The research data supporting this publication can be accessed at~\cite{suppdata}.
\end{acknowledgments}

\end{document}